\documentclass[onecolumn,secnumarabic,amssymb, amsmath, nofootinbib,tightenlines,
nobibnotes, aps, prl,epsfig]{revtex4}
\usepackage{graphicx}% Include figure files
\usepackage{dcolumn}% Align table columns on decimal point
\usepackage{bm}% bold math

\usepackage{amssymb}
\usepackage{epsfig}
\usepackage{color}

\newcommand{\ba}{\begin{eqnarray}}
\newcommand{\ea}{\end{eqnarray}}
\newcommand{\be}{\begin{equation}}
\newcommand{\ee}{\end{equation}}
\newcommand{\bdisplay}{\begin{displaymath}}
\newcommand{\edisplay}{\end{displaymath}}

\begin{document}
\preprint{APS/123-QED}
\title{The EIC reduced cross sections at high inelasticity}% Force line breaks with \\

\author{G. R. Boroun}%
 \email{ boroun@razi.ac.ir }
\affiliation{Department of physics, Razi University, Kermanshah
67149, Iran}% \textbackslash\textbackslash
%\author{B.Rezaei}%
% \email{ brezaei@razi.ac.ir }
%\affiliation{Department of physics, Razi University, Kermanshah
%67149, Iran}% \textbackslash\textbackslash

\date{\today}% It is always \today, today,

 \pacs{***}%PACS, the Physics and Astronomy
                              %Classification Scheme.
 %Use showkeys class option if keyword
\begin{abstract}
 The impact of  higher-twist corrections to the ratio $\frac{\sigma_{r}}{F_{2}}(A,Q^2/s,Q^2)$
 for light and heavy nuclei is considered at fixed a $\sqrt{s}$ and $Q^2$ to the minimum value of
$x$ given by $Q^2/s$. The results are for the EIC
center-of-mass energies. We apply the influence of higher-twist
corrections to $\frac{\sigma_{r}}{F_{2}}(A,Q^2/s,Q^2)$ by the
dimensionless variable $\xi'_{A}=Q_{s,A}^{2}/Q^2$ in the color
dipole model and obtain bounds for the ratio in the linear
region. The importance of contributing to twist-4 at small-$x$ is
visible in the linear region where $\xi'_{A}{\leq}1$. We perform
that twist-4 corrections of the saturation model are shown the
successful behavior of $\frac{F_{L}}{F_{2}}(A=2,x,Q^2)$ in
comparison with the JLab data at $\xi'_{A}{\leq}1$. The
higher-twist corrections due to twist-6 and twist-8 are resummed by
the non-linear approaches in the region $\xi'_{A}>1$. The ratio
$\frac{F_{L}}{F_{2}}$ deuteron is considered in the low-$Q^2$ and
small-$x$ region and compared with the JLab data which takes into
account the non-linear corrections due to  twist-6 and twist-8.
A comparison with the JLab data shows that the impact of the
distinct twists allows us to probe the presence of non-linear
effects on the QCD dynamics as $F_{L}{\rightarrow}0$ and the
polarization of the virtual photon is transverse in this
region. \\

%%%%%%%%%%%%%%%%%%%%%%%%%%%%%%%%%%%%%%%%%%%%%%%%%%%%%%%
\end{abstract}
 \pacs{***}%PACS, the Physics and Astronomy
                              %Classification Scheme.
 %Use showkeys class option if keyword
\maketitle
%**********************************************************
%%%%%%%%%%%%%%%%%%%%%%%%%%%%%%%%%%%%%%%%%%%%%%%%%%%%%%%%%%%%%%%%%%%%%%%%%%%%%%%%%%%%%%%%%%%%%
\section{I. Introduction}

The study of nuclear effects on the structure functions measured
in deep inelastic scattering (DIS) experiments at high energies is
one of the main goals of high energy colliders and offers
valuable information for understanding the dynamics of partons in
the nuclear environment \cite{EIC}. In the collinear factorized
approach to pQCD, the partonic structure of bound nuclei is
described by nuclear parton distribution functions (nPDFs)
\cite{Armesto, Asch, Armesto2}. nPDFs are interesting in
the transition between linear and non-linear (e.g., saturation) scale
evolution of the parton densities \cite{Marquet}. Saturation
occurs at low $x$ and low interaction scale $Q^2$ where the
recombination of low-$x$ gluons becomes increasingly important.
The transition line between the linear and non-linear regimes of
the QCD dynamics is described by the saturation scale $Q_{s}$,
which is predicted to depend on $x$ and atomic number A.
Increasing $Q_{s}$ at smaller values of $x$ and larger values of
A will be a scenario to determine whether parton distributions
saturate or not and allow us to disentangle non-linear from linear
physics.\\
 Taking into
account non-linear effects that are inherent to the QCD dynamics
at high energies when the hadron becomes a dense system can be
quite well described by models based on Color Glass Condensate
(CGC) formalism \cite{CGC}. The nuclear saturation effect
demonstrates that at small values of $x$, the gluon distribution
in a nucleus is less than the gluon distribution in a nucleon.
Such non-linearities are predicted to be more pronounced in
lepton-nucleus than in lepton-proton scattering which is the key
physics goal of an Electron-Ion Collider (EIC). Nuclear physics
with electron-nucleus (eA) collisions can be explored at the EIC,
where the maximum energy envisioned for electron-heavy ion runs
would be achieved by colliding $18~\mathrm{GeV}$ electrons with
$110~\mathrm{GeV}$ ions for a $\sqrt{s}=89~\mathrm{GeV}$ in the
EIC Conceptual Design Report \cite{EIC2}. Indeed, the EIC, which is
expected to begin science operations at Brookhaven National
Laboratory in the early 2030s, promises to revolutionize our
understanding of the internal structure of hadrons. The
longitudinal proton structure function $F_{L}$ at EIC through a
Rosenbluth separation method with impacts of differing assumptions
on sample sizes, systematic uncertainties and beam energy
scenarios are investigated in Refs.\cite{Lopez, LHeC}. In
Ref.\cite{Farid} a factorized expression of the differential
cross-section for single-inclusive jet production, for a
longitudinally polarized photon, in terms of transverse momentum
dependent (TMD) quark and gluon fracture functions is derived and
shown that the quark TMD fracture function is the most sensitive
to saturation effects in large nuclei.\\
The longitudinal structure function behavior on photon virtuality
$Q^2$ at fixed energy within the color dipole formalism for models
considering parton saturation effects which resum a wide class of
higher-twist (HT) contributions is considered in
Ref.\cite{Machado} based on the gluonic structure. The gluonic
structure of light nuclei that exclusive vector meson production
can provide, as well as the $x$ dependence of coherent and incoherent
cross sections at
the EIC are considered in \cite{Hiki}.\\
In the Operator Product Expansion (OPE), the scattering amplitudes
are expanded into a series of contributions
$\sigma=\sum_{\tau}\sigma_{\tau}(Q^2)$, with $Q^2$ being the
characteristic hard scale, $\sigma_{\tau}{\propto}1/Q^{\tau}$ and
$\tau=2,4,6$ and 8 being the twist\footnote{The notion of twist
was introduced in 1971 in the paper by Gross and Treiman
\cite{Twist2} who noticed that "it is no longer the dimension
alone that determines the importance of an operator near the
light-cone, but rather the difference between the dimension and
spin". They called this quantity the "twist" on an operator $\tau$
with the original definition $\tau$= dimension - spin is sometimes
referred to as "geometric twist". \cite{Twist3}.} \cite{Twist}.
Indeed, the higher twist corrections refer to a certain class of
contributions to hard processes in strong interactions that are
suppressed by a power of the hard scale. The influence of higher
twist corrections to deep inelastic structure functions in the
low-$Q^2$ and small-$x$ HERA region is investigated in
Ref.\cite{Bartels}. A twist analysis of the saturation model
describing the structure function and the DIS diffractive cross
section at HERA is performed in \cite{Bartels}. In Ref.\cite{Yan},
the impact of the higher-twist effects resummed by the non-linear
approaches for the QCD dynamics on the inclusive observable in EIC
is investigated.\\
In this paper, the behavior of the ratio
$\frac{\sigma_{r}}{F_{2}}(A,x_{\mathrm{min}},Q^2)$ due to the HT
corrections at  high inelasticity $y=1$ in the EIC COM energy
in a wide range of the mass number A is considered. In
Refs.\cite{Taylor, Boroun1n}, the behavior of the DIS structure
functions at high inelasticity, which is considered at the minimum
value of $x$ given by $Q^2/s$ where the polarization of the
exchanged photon is described to be transverse
and the longitudinal structure function is expected to be small at this kinematic domain.\\

\section{II. Method}

In the saturation model, the $\gamma^{*}A-\mathrm{cross}~
\mathrm{sections}$ are defined in terms of the dipole cross section
$\widehat{\sigma}(x,r^2)$ which describes the interaction of the
$q\overline{q}$ dipole pair with the nuclei by the gluonic field,
as well as the transverse and longitudinally polarized photon wave
functions $\Psi_{T,L}(z,r)$ by the following formula:
  \ba
\label{SigmaAD_eq} \sigma_{T,L}(A,x,Q^2)&=&\int
d^2{\mathbf{r}}\int dz \int d^{2}\mathbf{b}
|\Psi_{T,L}(z,\mathbf{r},Q^{2})|^2\widehat{\sigma}(x,r^2)S(\mathbf{b}),
 \ea
 where %
  \ba
\label{SigmaTL_eq}
F_{T,L}(A,x,Q^2)=&=&\frac{Q^2}{4\pi^2\alpha_{em}}\sigma_{T,L}(A,x,Q^2).
 \ea
The dipole cross section grows quadratically at small $r$ and 
saturates at large $r$, where $r$ is the relative transverse
separation between quark  and anti-quark. Here, $z$ is the
momentum fraction of the photon carried by the quark or anti-quark
in the color dipole and $\mathbf{b}$ is the transverse distance
from the center of the nucleus to the center of mass of the
$q\overline{q}$ dipole \cite{Nikolaev1}. $S(\mathbf{b})$ is the
profile function in impact parameter space,  usually
described by a Wood-Saxon distribution. The Mellin transform of
the wave functions from the dipole cross section to
evaluate the cross sections in Eq.~(\ref{SigmaAD_eq}) is conducted in
Ref.\cite{Bartels}.  The transverse cross section has a double
pole that generates a logarithmic behavior for the leading-twist
contribution, while the longitudinal leading-twist contribution
has only a single pole. Higher-twist contributions are obtained by
evaluating the
residues at the lower lying poles in \cite{Bartels}.\\
The transverse and longitudinal structure functions are calculated
\cite{Bartels, Yan} by  expanding in powers of
$\xi_{A}{\equiv}Q^{2}_{s,A}/Q^2$,  where $Q_{s,A}$  represents the saturation scale
for the nucleus. It is defined as
$Q^{2}_{s,A}=A^{1/3}Q_{0}^{2}(x_{0}/x)^\lambda$ with
$Q_{0}^{2}=1~\mathrm{GeV}^2$.  In electron-ion collisions, the
 target area and the transverse size of the dipole cross
section scaline with
$S{\rightarrow}S_{A}=A^{2/3}S$. The transverse and longitudinal
structure functions at a fixed $\sqrt{s}$ and $Q^2$ at
$x_{\mathrm{min}}=Q^2/s$ can be expressed in terms of the parameter
$\xi_{A}$
 modified as $\xi_{A}{\rightarrow}\xi'_{A}=A^{1/3}\frac{Q_{0}^{2}}{Q^2}(\frac{sx_{0}}{Q^2})^\lambda
 $. This limit is crucial as it describes  the polarization of the exchanged photon as transverse, with the longitudinal structure function expected to be small.
At low values of $x$, the dominant gluon component in the
longitudinal structure function is strongly suppressed
\cite{Taylor, Boroun1n}. Therefore, we have
  \ba
\label{FLnA_eq}
 F_{k}^{\tau}(A,Q^2/s,Q^2)&=&\frac{Q^2}{4{\pi^2}\alpha_{em}}A^{2/3}\sigma_{0}\sum{e_{f}^{2}}\frac{\alpha_{em}}{\pi}\sum_{n=2}\eta^{k}_{n}{\xi'}_{A}^{n/2}~~~ \mathrm{with}~ n=\tau~
 \mathrm{even}~ \mathrm{and}~~ k=T,L,
 \ea
where $n$ indicates the twist-2, 4, 6 and 8 analysis. The
coefficient functions $\eta_{n}$ according to Refs.\cite{Bartels,
Yan} are:
  \ba
\label{AnA_eq}
 \eta^{L}_{2}&=&1 \nonumber\\
  \eta^{T}_{2}&=&\frac{7}{6}-\psi(2)+{\ln}(\frac{1}{{\xi'}_{A}}) \nonumber\\
 \eta^{L}_{4}&=&-\frac{94}{75}+\frac{4}{5}\psi(3)-\frac{4}{5}{\ln}(\frac{1}{{\xi'}_{A}})
 \nonumber\\
 \eta^{T}_{4}&=&\frac{6}{10}
 \nonumber\\
\eta^{L}_{6}&=&\frac{654}{1225}-\frac{36}{35}\psi(4)+\frac{36}{35}{\ln}(\frac{1}{{\xi'}_{A}})
 \nonumber\\
 \eta^{T}_{6}&=&\frac{43}{1225}-\frac{12}{35}\psi(4)+\frac{12}{35}{\ln}(\frac{1}{{\xi'}_{A}})
 \nonumber\\
 \eta^{L}_{8}&=&-\frac{1636}{18375}+\frac{48}{175}\psi(5)-\frac{48}{175}{\ln}(\frac{1}{{\xi'}_{A}})\nonumber\\
 \eta^{T}_{8}&=&-\frac{262}{11025}+\frac{4}{35}\psi(5)-\frac{4}{35}{\ln}(\frac{1}{{\xi'}_{A}}).
 \ea
The inclusive $eA$ scattering cross- section at low
$Q^2$ and $x$ at EIC will be bounded due to kinematic
factors. The cross- section in reduced form is given by
  \ba
\label{ReducA_eq}
 \frac{\sigma_{r}}{F_{2}}(A,Q^2/s,Q^2)&=&1-\frac{F_{L}}{F_{2}}(A,Q^2/s,Q^2)=1-\frac{F_{L}}{F_{T}+F_{L}}(A,Q^2/s,Q^2)=
 \left[1+\frac{F_{L}^{\tau}}{F_{T}^{\tau}}(A,Q^2/s,Q^2)\right]^{-1}\nonumber\\
 &&=\left[1+\frac{\sum_{\tau}\eta^{L}_{\tau}{\xi'}_{A}^{n/2}}{\sum_{\tau}\eta^{T}_{\tau}{\xi'}_{A}^{n/2}}\right]^{-1}.
 \ea
The dipole cross- section depends on  $x$ and $r$ according to \cite{Golec1, Golec2}
  \ba
\label{SigmaADipole_eq} \int d^{2}\mathbf{b}
\widehat{\sigma}(x,r^2)S(\mathbf{b})&=&2[1-\exp(-r^2Q^{2}_{s,A}/4)]\int
d^{2}\mathbf{b}
S(\mathbf{b})=A^{2/3}\sigma_{0}[1-\exp(-r^2Q^{2}_{s,A}/4)].
 \ea
In the color dipole model (CDM), the cross section in reduced form
is:
  \ba
\label{ReducACDM_eq}
 \frac{\sigma_{r}}{F_{2}}(A,Q^2/s,Q^2)&=&1-\frac{\int
d^2{\mathbf{r}}\int dz
|\Psi_{L}(z,\mathbf{r},Q^{2})|^2[1-\exp(-r^2Q^{'2}_{s,A}/4)]}{\int
d^2{\mathbf{r}}\int dz
\{|\Psi_{T}(z,\mathbf{r},Q^{2})|^2+|\Psi_{L}(z,\mathbf{r},Q^{2})|^2\}[1-\exp(-r^2Q^{'2}_{s,A}/4)]},
 \ea
where $Q^{'2}_{s,A}=A^{1/3}Q_{0}^{2}(sx_{0}/Q^2)^\lambda$. In
Eq.~(\ref{ReducACDM_eq}), $\Psi_{L,T}$ represent the spin-
averaged light-cone wave functions of the photon. The square of
the photon wave function indicates the probability  of a $(q\overline{q})$ fluctuation occurring with a transverse size
relative to the photon polarization. The light-cone photon
wave function, $\Psi$, is represented  by the lowest order
$\gamma^{*}g{\rightarrow}q\overline{q}$ scattering amplitudes
which give
  \ba
\label{PsiT_eq}
|\Psi_{T}^{f}(z,r,Q^2)|^2=\frac{2N_{c}\alpha_{em}e_{f}^{2}}{4\pi^2}\bigg{\{}
[z^2+(1-z)^2]\epsilon^2K_{1}^{2}(\epsilon{r})+m_{f}^{2}K_{0}^{2}(\epsilon{r})\bigg{\}}
\ea
and
  \ba
\label{PsiL_eq}
|\Psi_{L}^{f}(z,r,Q^2)|^2=\frac{8N_{c}\alpha_{em}e_{f}^{2}}{4\pi^2}
Q^2z^2(1-z)^2K_{0}^{2}(\epsilon{r}),
 \ea
where $\epsilon^2=z(1-z)Q^2+m_{f}^{2}$, $K_{0}$ and $K_{1}$ are
modified Bessel functions, and the sum is over quark flavors $f$
with quark mass $m_{f}$.\\
In the following the validity of the higher-twist corrections to
the nuclear reduced cross section with respect to the EIC COM
energy  at the critical points between the linear and non-linear
regimes according to the high inelasticity due to the $\xi'_{A}$ variable, will be investigated.\\

\section{III. Results and conclusion}

The transition point between the linear and non-linear regimes occurs
when $\xi'_{A}{\approx}1$ and is proportional to the mass number A
and the $Q^2$ variable at a fixed $\sqrt{s}$ and high inelasticity.
The perturbative region is proportional to the linear region with
$\xi'_{A}{\lesssim}1$ at  large $Q^2$.  At small $Q^2$, the
nonperturbative region with $\xi'_{A}{>}1$ is proportional to the
non-linear region defined by the Pomeron region
\cite{Bartels}. In Fig.1  the linear and non-linear regions are shown due to
the large and small values of $\xi'_{A}$ at the limit
$x_{\mathrm{min}}=Q^2/s$ according to the EIC collider with COM
energy of $\sqrt{s}=89~\mathrm{GeV}$ for $n_{f}=3$ is shown. The
theoretical expectations for the transition point are obtained using
the parameters $Q_{0}^{2}=1~\mathrm{GeV}^2$,
$\sigma_{0}=23.58~\mathrm{mb}$, $\lambda=0.270$,
$x_{0}=2.24{\times}10^{-4}$, and $m_{f}=0.14~\mathrm{GeV}$
\cite{Golec3} in Fig.1. The critical point increases towards large
$Q^2$ values as the mass number A increases. These critical
points, which indicate the boundary between the linear and
non-linear regimes, are  approximately obtained for Deuterium-2,
Carbon-12, Fe-56, and Lead-208  at the $Q^2$ values
$Q^2{\lesssim}2$, $3$, $4$, and $6~\mathrm{GeV}^2$.\\
\begin{figure}
\centering
\includegraphics[width=0.55\textwidth]{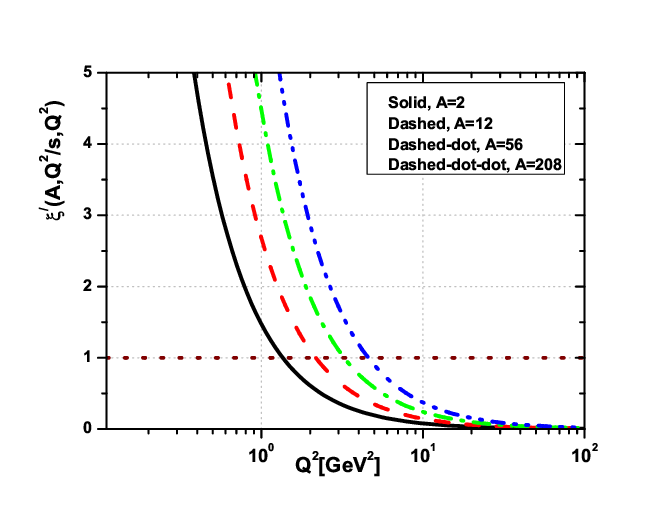}
\caption{Linear ($\xi'_{A}{\lesssim}1$) and non-linear
($\xi'_{A}{>}1$) regions are
plotted as a function of $Q^2$ and nuclear mass number A for $n_{f}=3$,  according to the transition point
$\xi'_{A}{\approx}1$ (short-dashed-brown line). The curves represent  
(Deuterium-2 (solid-black curve), Carbon-12 (dashed-red curve),
Fe-56 (dashed-dot-green curve), and Lead-208 (dashed-dot-dot-blue
curve) at the EIC COM energy with
$\sqrt{s}=89~\mathrm{GeV}$.}\label{Fig1}
\end{figure}
\begin{figure}
\centering
\includegraphics[width=0.9\textwidth]{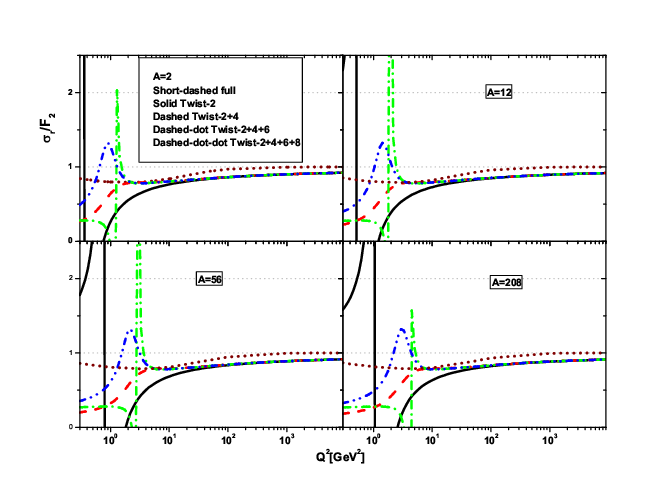}
\caption{ The higher-twist corrections, including  twist-2 (solid-black
curve), twist-2+4 (dashed-red curve), twist-2+4+6
(dashed-dot-green curve), and twist-2+4+6+8 (dashed-dot-dot-blue
curve), to the ratio $\frac{\sigma_{r}}{F_{2}}$ at the EIC COM
energy with $\sqrt{s}=89~\mathrm{GeV}$ with
$x_{\mathrm{min}}=Q^2/s$ are compared with the GBW model
(short-dashed-brown full curve) for $n_{f}=3$ and nuclei
Deuterium-2, Carbon-12, Fe-56, and Lead-208.}\label{Fig1}
\end{figure}
In Fig.2, the contributions of  distinct twists to the ratio
$\frac{\sigma_{r}}{F_{2}}(A,Q^2/s,Q^2)$ for various nuclei across a
wide  range of $Q^2$ values  at the EIC COM energy with
$\sqrt{s}=89~\mathrm{GeV}$ and $x_{\mathrm{min}}=Q^2/s$  for
$n_{f}=3 $ are analyzed and compared with the GBW model. It is
observed that the distinct twists exhibit similar behavior at large
values of $Q^2$, with the ratio $\frac{\sigma_{r}}{F_{2}}<1$ at this
limit being independent of the twists and the increasing 
mass number A. The full results with the GBW model  show that
the ratio $\frac{\sigma_{r}}{F_{2}}(A,Q^2/s,Q^2)=1$ at large
values of $Q^2$. The intersection point between the GBW model and
twist-2+4  indicates the critical point between the linear and
non-linear regimes in Fig.2. This point shifts towards larger
$Q^2$ values as the mass number A increases. The effects of
twist-6 and twist-8 at low $Q^2$ values are more noticeable. The
ratio $\frac{\sigma_{r}}{F_{2}}{\rightarrow}1$ at low $Q^2$ values
in the GBW model and higher twist corrections, as the longitudinal
structure function is approximately $F_{L}{\rightarrow}0$ as
$Q^2{\rightarrow}0$,  indicating that the polarization of the
exchanged photon is transverse. The ratio from the GBW model is
larger than the higher twist corrections at both small and large values
of $Q^2$.\\
The impact of higher twist corrections on
$\frac{\sigma_{r}}{F_{2}}(A,Q^2/s,Q^2)$ is minimal at large $Q^2$ values, which also holds true for larger values of
A. However,  the main difference is that the impact of higher twist
corrections becomes significant at small values of
$Q^2$. The ratio
 is significantly influenced by the twist-4+6+8 terms, which provide substantial positive corrections to the leading-twist contribution
 at low $Q^2$. The disparity between the higher twist corrections
 (i.e., twist-2, twist-2+4, twist-2+4+6 and twist-2+4+6+8) and
the full predictions from the GBW model for the ratio $\frac{\sigma_{r}}{F_{2}}(A,Q^2/s,Q^2)$
 diminishes to less than $8.8\%$ with Lead-208 and $7.7\%$ with
 Deuterium-2 for large $Q^2$ values.\\
For comparison with experimental data, we calculated the ratio
$\frac{F_{L}}{F_{2}}(A,x,Q^2)$ for  deuterium and compared it with
the results obtained at Jefferson Lab (JLab) \cite{JLAB} in
Fig.3. The error bars of the ratio
$F_{L2}{\equiv}\frac{F_{L}}{F_{2}}$ in the JLab data are
determined by the following formula:
 $$\Delta({F_{L2}})=F_{L2}\sqrt{(\Delta{F_{L}}/F_{L})^2+(\Delta{F_{2}}/F_{2})^2},$$
where in the JLab data, $\Delta{F_{L}}$ and  $\Delta{F_{2}}$ are
taken from the JLab E00-002 \cite{JLAB}. The parameter
$\xi'_{A}$ for the results of the ratio
$\frac{F_{L}}{F_{2}}(A=2,x,Q^2)$ indicates that the data from JLab
are in the linear region when $\xi'_{A}\leq 1$ for $x{>}0.02$ at
$Q^2=0.4~\mathrm{GeV}^2$. In Fig.3, the ratio
$\frac{F_{L}}{F_{2}}$ for  deuterium is determined at
$Q^2=0.4~\mathrm{GeV}^2$ for $n_{f}=3$ to account for higher twist
corrections and compared  with the results from JLab and
the CDM\footnote{For further discussion, refer to Appendix A.}. We
observe that the data in the linear regime with twists-(2+4, 2+4+6
and 2+4+6+8) are comparable to the JLab data, showing a
significant negative correction to the leading twist contribution.
The JLab data in Fig.3 approaches
$\frac{F_{L}}{F_{2}}{\rightarrow}0$ as the higher twist
corrections from twist-6 and twist-8 at low $Q^2$ values
indicate that the polarization of the virtual photon is
transverse when the longitudinal structure function approaches
 low $x$ values. We conclude that the results at low $Q^2$
values with twist-4 strongly inflience the results in the linear
region, valid for $\xi'_{A}{\lesssim}1$, while twist-6 and
twist-8  affect the saturation region where
$Q_{s,A}^2{\geq}Q^2$ and non-linear effects amplify and
modify the observables at low $Q^2$ and $x$ values in the EIC.\\

\begin{figure}
\centering
\includegraphics[width=.55\textwidth]{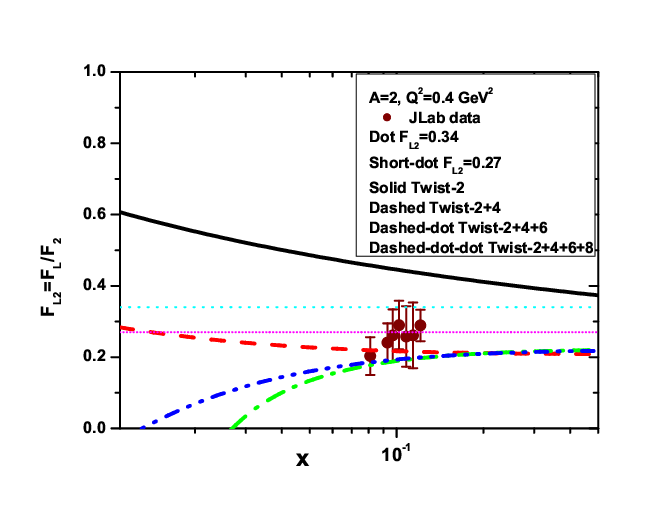}
\caption{The extracted ratio $F_{L2}$ of Deuterium in the twist
expansion for twist2 (solid-black curve), twist-2+4 (dashed-red
curve), twist-2+4+6 (dashed-dot-green curve) and twist-2+4+6+8
(dashed-dot-dot-blue curve) as a function of $x$ is plotted at
$Q^2=0.4~\mathrm{GeV}^2$ at the active flavor numbers $n_{f}=3$
and compared with the data from JLab E00-002 \cite{JLAB} (shown as
brown points). The results are compared to the
 dipole upper bounds  by
lines corresponding to $F_{L2}=0.27$ (short-dot-magenta line) and
$F_{L2}=0.34$ (dot-cyan line).}\label{Fig8}
\end{figure}

In conclusion, we have examined the behavior of the ratio
$\frac{\sigma_{r}}{F_{2}}(A,Q^2/s,Q^2)$ for light and heavy nuclei
at $y=1$, where $x_{\mathrm{Bj}}=x_{\mathrm{min}}=Q^2/s$, in the
EIC kinematics. We utilized the CDM and the twist expansion for
low $x$ in this scenario. Our model provides a good description of
results at moderate and large $Q^2$ values and predicts nuclear
ratio $\frac{\sigma_{r}}{F_{2}}(A,Q^2/s,Q^2)$  that can be
measured in electron-ion collisions at $x_{\mathrm{min}}$. These
results at moderate and large $Q^2$ values strongly suggests that
the gluon component is the dominant factor in the longitudinal
structure function within the dipole picture. We observed that the
value of $F_{L}$ is small at low $Q^2$ values in the dipole
picture. This is because the dominant gluon component is strongly
suppressed and the polarization of the exchanged photon is
transverse at this kinematic point. We investigated the nuclear
ratio $\frac{\sigma_{r}}{F_{2}}(A,Q^2/s,Q^2)$ at this limit, which
reveals high gluon densities and associated nonlinear high-energy
evolution. The non-linear corrections at the saturation regime
increase with the mass number $A$. The ratio
$\frac{F^{A}_{L}}{F^{A}_{2}}$ can serve as an indicator for the
presence of non-linear low $x$ dynamics in large nuclei. The ratio
of $F_{L2}$ for deuterium at low four-momentum transfer squared
is determined and compared with the JLab E00-002 data and the CDM
bounds. Indeed, the search for non-linear effects in light and
heavy nuclei based on the EIC kinematics is investigated because
the non-linear regime is enhanced by a factor ${\propto}A^{1/3}$, the
nuclear saturation scale as this approaches resums higher twist
contributions at low $Q^2$ as observed in Refs.\cite{Bartels, Yan}.\\

%\subsection{ACKNOWLEDGMENTS}

%The author would like to thank  F.E.Taylor and F.Salazar for their
%helpful comments and invaluable support. The author is especially
%grateful to Nestor Armesto, Elke Aschenauer, and Paul Newman for
% fruitful discussions and for allowing access to data
%related to the longitudinal structure function at the EIC and
%JLab. I am also very grateful to the Department of Physics at
%CERN-TH for their
%warm hospitality.\\
%%%%%%%%%%%%%%%%%%%%%%%%%%%%%%%%%%%%%%%
\section{Appendix A}

In Refs.\cite{Ref14, Ref15, Ref17}, the authors demonstrate that at large
$Q^2$, the ratio of photo absorption cross sections
 is determined by a parameter $\rho$, which describes the dissociation
 of photons into $q\bar{q}$ pairs, $\gamma^*_{L,T}\rightarrow q\bar{q}$, with the realtion:
  \ba
\label{Rho_eq}
 R=\frac{1}{2\rho}, \ea
where the factor 2 in the equation originates from the difference in the photon
wave functions. Essentially, the $\rho$ parameter represents the ratio
of the average transverse momenta, given by:
$\rho=\frac{<\overrightarrow{k}^{2}_{\bot}>_{L}}{<\overrightarrow{k}^{2}_{\bot}>_{T}}$,
or it can  be related to the ratio of the effective transverse
sizes of the $(q\overline{q})^{J=1}_{L,T}$ states as
$\frac{<\overrightarrow{r}^{2}_{\bot}>_{L}}{<\overrightarrow{r}^{2}_{\bot}>_{T}}=\frac{1}{\rho}$.
For the specific value of $\rho=\frac{4}{3}$ in Ref.\cite{Ref15}
\footnote{For further discussion, readers can refer to the papers
by M. Kuroda and D. Schildknecht: "Phys. Rev. D 85  (2012) 094001"
and
 "J. Mod. Phys. A 31 (2016)
1650157". }, it is found that $R=\frac{3}{8}=0.375$ and the ratio
of structure functions is $F_{L2}(x,Q^2)=\frac{3}{11}=0.273$. When $\rho=1$ (i.e., helicity independent), the
ratio is  approximately $0.34$, serving as an upper bound for
$F_{L2}(x,Q^2)$  in the dipole model. In Refs.\cite{Ref18, Ref19,
Ref20}, the authors establish that the ratio of structure functions in the
dipole model is independent of the dipole cross section
$\sigma_{\mathrm{dip}}$, being proportional to the
photon-$q\overline{q}$ wave function as
  \ba
\label{FL2_eq} F_{L2}(x,Q^2)=g(Q,r,m_{q}) {\leq}
\widetilde{g}(z_{m})=0.27139, \ea
where
  \ba
\label{gL2_eq} g(Q,r,m_{q})=\frac{ \int_{0}^{1} dz
|\Psi_{L}(\mathbf{r},z;Q^{2})|^{2}}{ \int_{0}^{1} dz
\bigg{[}|\Psi_{T}(\mathbf{r},z;Q^{2})|^{2}+|\Psi_{L}(\mathbf{r},z;Q^{2})|^{2}\bigg{]}}
\ea
and $m_{q}$ represents the mass of the active quark \footnote{For further
discussion see \cite{Ref19}.}. For massless quarks, the function
$g(Q,r,m_{q})$ is defined by the dimensionless variable $z=Qr$, where
the function $\widetilde{g}(z)=g(Q,r,0)$ reaches its maximum at
$z_{m}=2.5915$ with $\widetilde{g}(z)=0.27139$. Previous studies \cite{Ref21, Ref22, Ref23, Ref24} have shown that the bound specified by
Eq.~(\ref{FL2_eq}) for the ratio of structure functions holds true for all
$Q{\geq}0$, $r{\geq}0$ and $m_{q}{\geq}0$.\\

%%%%%%%%%%%%%%%%%%%%%%%%%%%%%%%%%%%%%%%%%%%%%%%%%%%%%%%%%%%%%%%%%%%%

\end{document}